# Analysis on the Foucault pendulum by De Alembert Principle and Numerical Simulation

Zhiwu Zheng

Department of Physics, Nanjing University, Jiangsu, China 210046

**ABSTRACT**

In this paper, we handle the problem of the motion of the Foucault pendulum. We explore a new method induced from the De Alembert Principle giving the motional equations without small-amplitude oscillation approximation. The result of the problem is illustrated by numerical analysis showing the non-linear features and then with a comparison with a common method, showing the merit of this new original method. The result also shows that the argument changes in near-harmonic mode and the swing plane changes in pulsing way.

I. INTRODUCTION

The Foucault pendulum usually consists of a large weight suspended on a cable attached to a point two or more floors above the weight. [1-5] As the pendulum swings back and forth its plane of oscillation rotates clockwise in the northern hemisphere, convincing the rotation of the Earth.

The motion of the Foucault pendulum is the coupling of two periodic motion, the rotation of the Earth and the oscillation of the pendulum leading to a precession [2-6]. Depending on the relevant parameters, the motion differs in varied conditions, showing periodic or non-periodic motion as a consequence. With gravity, tension and Corioli force applied on the cable, the motional equation, due to Newton's Law, is difficult in solution.

Up to now, a lot of researches were focused on the Foucault pendulum, and to be special, in some references [7-10], these authors gave assumption that the weight could be seen as moving in the horizontal plane and not moving in the vertical direction due to its low-amplitude oscillation and propose the motional equations of the pendulum which can get analytical answer to them. The equations can be expressed as:

$$\ddot{\alpha} - \alpha \dot{\beta}^2 - 2w \sin\varphi \, \alpha\dot{\beta} + \frac{g}{l} \alpha = 0 \tag{0.1}$$

$$\alpha \ddot{\beta} + 2 \dot{\alpha}\dot{\beta} + 2w \sin\varphi \, \dot{\alpha} = 0 \tag{0.2}$$

The angles, α, β, (afterwards shown in the figure 2) are the intersection angles between z axis and the cable, x axis and the projection on x-y plane respectively. And $\varphi$ is the latitude locating the pendulum, $w$ is the angular velocity of the earth rotation.

The equation (0.2) can be expressed as follow by integration

$$\alpha^2 \dot{\beta} + 2w \sin\varphi \, \alpha^2 = \text{Constant} \tag{0.3}$$

These equations, (0.1) and (0.3), with their low-amplitude oscillation approximation, can be solved analytically in special initial condition.

However, we can get that the assumption in those references stating that the weight moves in a 2-dimensional way is unconvincing in terms of not only high-amplitude oscillation but also the long-time process through the research described afterwards, which is to analyze the system firstly without the small-angle approximation. And we have the necessity to reanalyze this problem and give the supplement of these conditions.

In this paper, what is mainly concentrated on is how the Foucault pendulum moves in varied relevant conditions. The model giving some assumptions of the pendulum itself and the property of the earth will be first built up and then with the theoretical result and some analysis. Finally, the numerical analysis and simulation will be used to analyze different motion types.

## II.   THEORETICAL MODEL

### 1. Assumptions

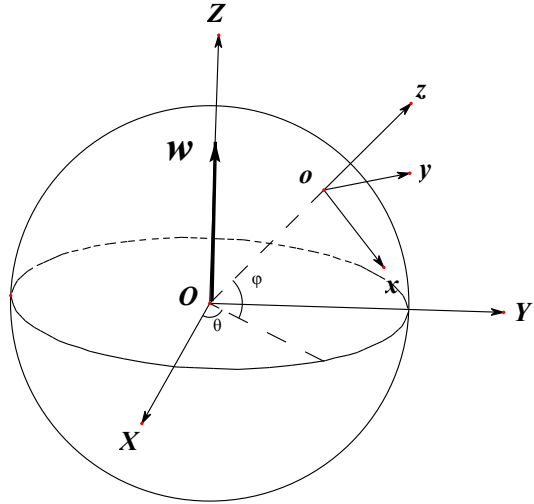

**Figure 1:**   The formation of coordinates of
   the earth and the pendulum

We first will give some reasonable assumptions of the Foucault pendulum we concentrated on:

a) Massless cable

   We consider a massless cable with its length $l$ tied between the weight and the fixed point.

b) Mass point for the weight

   The weight is treated as a mass point ignoring its influence of size and shape with mass $m$.

c) Negligible motion range

   The range of the motion can be neglected compared with the size of the earth. Therefore, we neglect the variation of the latitude and the gravitational field of the weight.

d) The shape and the rotation of the earth

   We consider the earth as a sphere, with its uniform gravitational field $g$ on its surface, rotating with a uniform angular velocity $w$ around its geometrical axis.

Other usual considerations of the motion of a particle in the Classic Mechanics need not be described here. [11-13]

## 2. The Construction of Coordinate System and Reference System

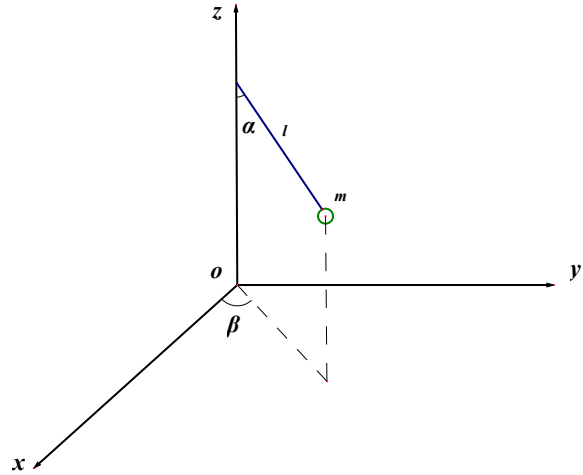

**Figure 2: The pendulum and the coordinate**

To deduce the motional equations of the pendulum, the coordinate system and reference system should first be given. And in order to give the motion feature of the pendulum, a coordinate system fixed on the point, with latitude $\varphi$ and longitude $\theta$, lying the pendulum and rotating with the earth firstly should be given, as the figure 1 shows, the Cartesian coordinates o-xyz, with its base vectors $\boldsymbol{i}, \boldsymbol{j}, \boldsymbol{k}$. So as to describe the angular velocity in this system, an assistant coordinate system should be given, as the figure 1 shows, the O-XYZ, with its base vectors $\boldsymbol{e_x}, \boldsymbol{e_y}, \boldsymbol{e_z}$. The relationship between this two systems is:

$$\boldsymbol{i} = sin\varphi\, cos\theta\, \boldsymbol{e_x} + sin\varphi\, sin\theta\, \boldsymbol{e_y} - cos\varphi\, \boldsymbol{e_z}$$

$$\boldsymbol{j} = -sin\theta\, \boldsymbol{e_x} + cos\theta\, \boldsymbol{e_y}$$

$$\boldsymbol{k} = cos\varphi\, cos\theta\, \boldsymbol{e_x} + cos\varphi\, sin\theta\, \boldsymbol{e_y} + sin\varphi\, \boldsymbol{e_z}$$

Additionally, the angular velocity is: $\boldsymbol{w} = w\, \boldsymbol{e_z}$. The components of $\boldsymbol{w}$ in the o-xyz system can be given as:

$$w_x = -w\, cos\varphi$$

$$w_y = 0$$

$$w_z = w\sin\varphi$$

(1)

3. The deformation of the motional equations

If, ordinarily, we use Newton's law to solve this problem, the effect of the tension of the cable must be taken into consideration. In order to simplify the process, to avoid the complicated effect of the tension of the rope, in this model, the De Alembert Principle will be used and the angles, α, β, shown in the figure 2, the intersection angles between z axis and the cable, x axis and the projection on x-y plane respectively, will be taken as generalized coordinates. (Indeed, the work of the tension equals zero when we take these generalized coordinates.)

The component of the position of the weight can be described with these generalized coordinates as:

$$x = l\sin\alpha\cos\beta$$
$$y = l\sin\alpha\sin\beta$$
$$z = h - l\cos\alpha$$

Thus, we can get:

$$\dot{x} = l\cos\alpha\cos\beta\,\dot{\alpha} - l\sin\alpha\sin\beta\,\dot{\beta}$$

$$\dot{y} = l\cos\alpha\sin\beta\,\dot{\alpha} + l\sin\alpha\cos\beta\,\dot{\beta}$$

$$\dot{z} = l\sin\alpha\,\dot{\alpha}$$

(2.1)

And

$$\delta x = l\cos\alpha\cos\beta\,\delta\alpha - l\sin\alpha\sin\beta\,\delta\beta$$

$$\delta y = l\cos\alpha\sin\beta\,\delta\alpha + l\sin\alpha\cos\beta\,\delta\beta$$

$$\delta z = l\sin\alpha\,\delta\alpha$$

(2.2)

$$\ddot{x} = l\cos\alpha\cos\beta\ddot\alpha - l\sin\alpha\sin\beta\ddot\beta - l\sin\alpha\cos\beta\dot\alpha^2 - 2l\cos\alpha\sin\beta\,\dot\alpha\dot\beta - l\sin\alpha\cos\beta\dot\beta^2$$
$$\ddot{y} = l\cos\alpha\sin\beta\ddot\alpha + l\sin\alpha\cos\beta\ddot\beta - l\sin\alpha\sin\beta\dot\alpha^2 + 2l\cos\alpha\cos\beta\,\dot\alpha\dot\beta - l\sin\alpha\sin\beta\dot\beta^2$$
$$\ddot{z} = l\sin\alpha\ddot\alpha + l\cos\alpha\dot\alpha^2$$

(2.3)

The applied force on the weight (having the real effect) is:
$$\boldsymbol{F} = \boldsymbol{F_C} + \boldsymbol{G}$$

Where
$$\boldsymbol{G} = -mg\boldsymbol{k}$$

$$\boldsymbol{F_C} = 2m\mathbf{v}\times\mathbf{w}$$

Where
$$\mathbf{v} = \dot{x}\boldsymbol{i} + \dot{y}\boldsymbol{j} + \dot{z}\boldsymbol{k}$$

$$\mathbf{w} = w_x\,\boldsymbol{i} + w_y\,\boldsymbol{j} + w_z\,\boldsymbol{k}$$

(3.1)

Due to the De Alembert Principle,

$$\delta W = (\boldsymbol{F} - m\ddot{\boldsymbol{r}})\cdot \delta \boldsymbol{r} = 0$$

And using the generalized coordinates, that is

$$\delta W = Q_\alpha\,\delta\alpha + Q_\beta\,\delta\beta$$

$Q_\alpha$  $Q_\beta$  are the generalized forces.

Get that:
$$Q_\alpha = 0$$
$$Q_\beta = 0$$

When substituted (1), (2.2), (2.3), and (3.1), we can deduce that:

$$\ddot\alpha - \sin\alpha\cos\alpha\,\dot\beta^2 - 2w\sin\alpha\,(\cos\varphi\sin\alpha\cos\beta + \sin\varphi\cos\alpha)\,\dot\beta + \frac{g}{l}\sin\alpha = 0$$

(3.2)

$$\sin\alpha\,\ddot\beta + 2\cos\alpha\,\dot\alpha\dot\beta + 2w\,(\cos\varphi\sin\alpha\cos\beta + \sin\varphi\cos\alpha)\dot\alpha = 0$$

(3.3)

From (3.2) and (3.3), we can theoretically get the motion features of the Foucault pendulum in the reference system rotating with the earth. To get special features, we need to solve these two differential equations.

Putting the two groups of equations above ((0.1), (0.2), and (3.2), (3.3)) together,

$$\ddot{\alpha} - \alpha\,\dot{\beta}^2 - 2w\,sin\varphi\,\alpha\dot{\beta} + \frac{g}{l}\,\alpha = 0$$

$$\alpha\,\ddot{\beta} + 2\,\dot{\alpha}\dot{\beta} + 2w\,sin\varphi\,\dot{\alpha} = 0$$

$$\ddot{\alpha} - sin\alpha\,cos\alpha\,\dot{\beta}^2 - 2wsin\alpha\,(cos\varphi\,sin\alpha\,cos\beta + sin\varphi\,cos\alpha)\,\dot{\beta} + \frac{g}{l}\,sin\alpha = 0$$

$$sin\alpha\,\ddot{\beta} + 2\,cos\alpha\,\dot{\alpha}\dot{\beta} + 2w\,(cos\varphi\,sin\alpha\,cos\beta + sin\varphi\,cos\alpha)\dot{\alpha} = 0$$

We can get that if putting $sin\alpha \approx \alpha$, $cos\alpha \approx 1$ due to small angle $\alpha$ and $cos\varphi\,sin\alpha\,cos\beta \approx 0$ compared with the other terms, these two groups of equations will be equivalent. When $\varphi = 0$, that is to locate the pendulum at the North Pole, in addition, these two groups of equations are of equivalence.

Then, the numerical analysis will be given as a result and a comparison.

III. NUMERICAL ANALYSIS

In this problem, pursuing to the result as accurate as possible, the Runge-Kutta method [14, 15] will be used. The latitude of the pendulum will be fixed at $\varphi = \frac{\pi}{6}$, the length of rope is $l = 67m$, and the angular velocity of the rotation of the earth can be calculated as $w = 7.27 \times 10^{-5}\ rad/s$.

Several plots will firstly be given, then with the analysis and the comparison.

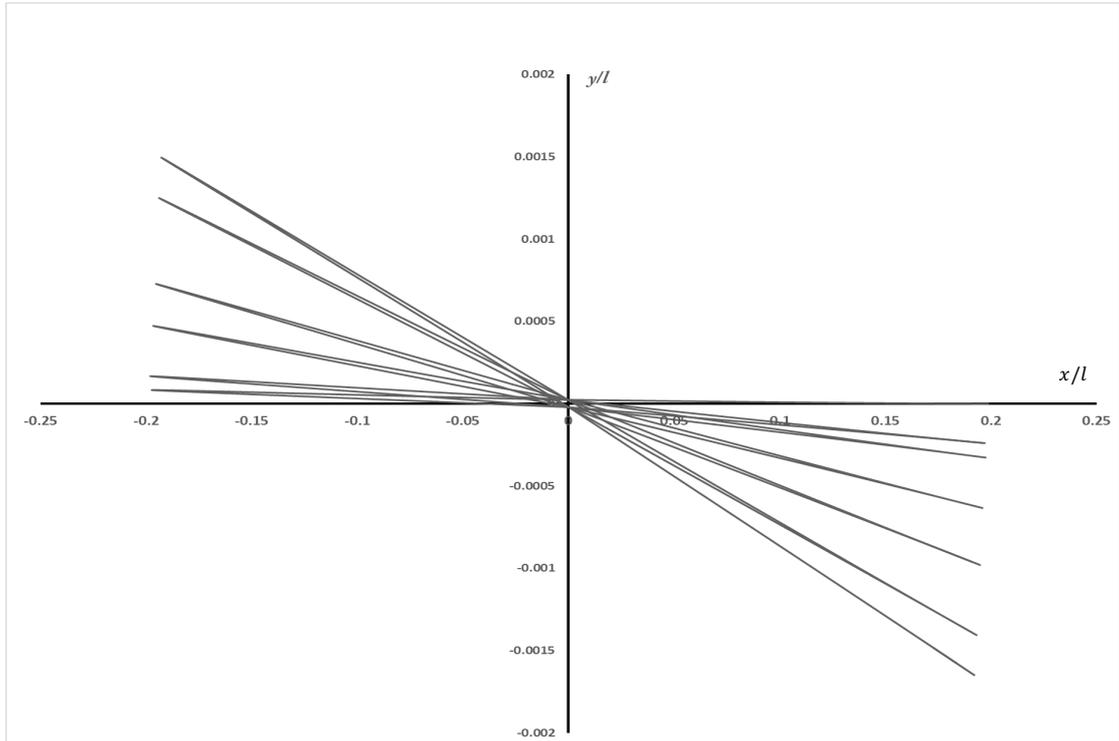

**Figure 3:** The motion trial projected on the horizontal plane for 100 s, derived from the new method (Equation A). Initial parameter: $l = 67m, \alpha_0 = 0.2 rad, \beta_0 = 0, \dot{\alpha}_0 = 0, \dot{\beta}_0 = 0$.

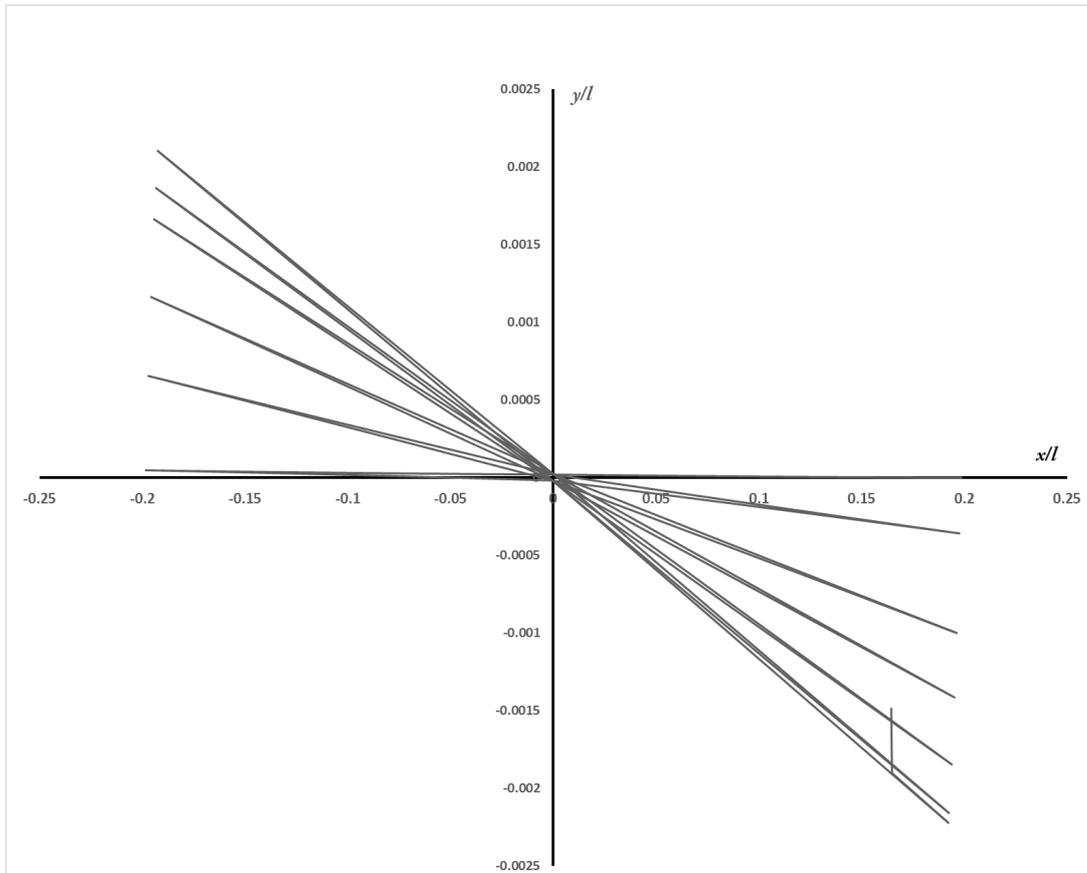

**Figure 4: The motion trial projected on the horizontal plane for 100 s, derived from the old method (Equation B). Initial parameter:** $l = 67m, \alpha_0 = 0.2 rad, \beta_0 = 0, \dot{\alpha}_0 = 0,$

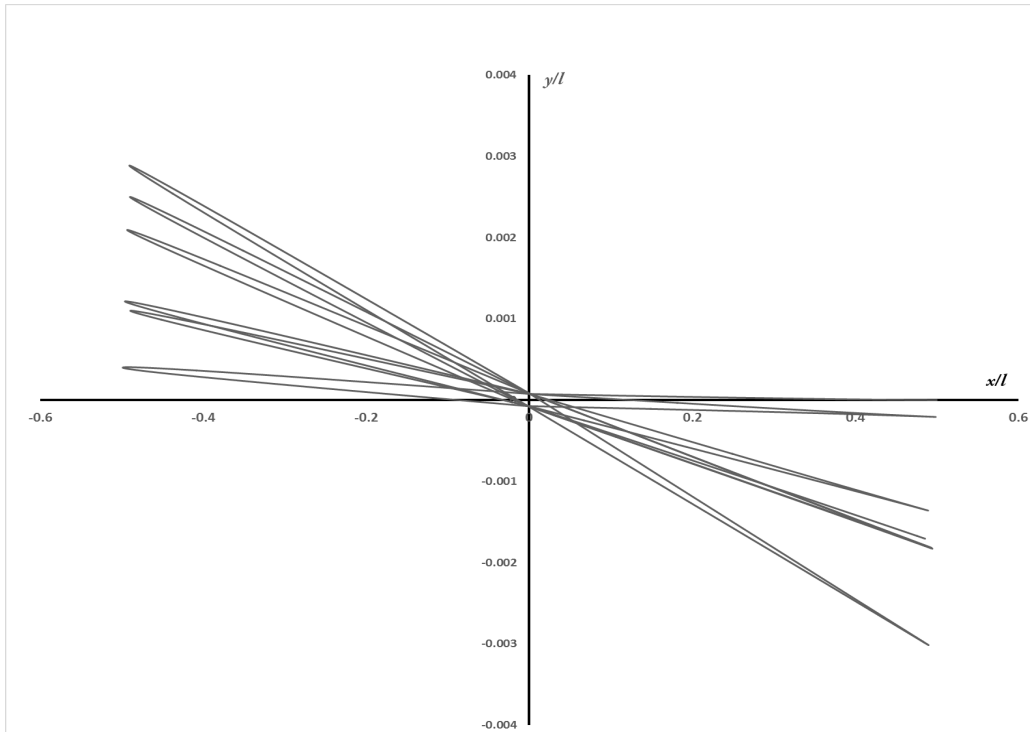

**Figure 5:** The motion trial projected on the horizontal plane for 100 s, derived from the new original method (Equation A). Initial parameter: $l = 67m, \alpha_0 = \pi/6, \beta_0 = 0, \dot{\alpha}_0 = 0, \dot{\beta}_0 = 0$.

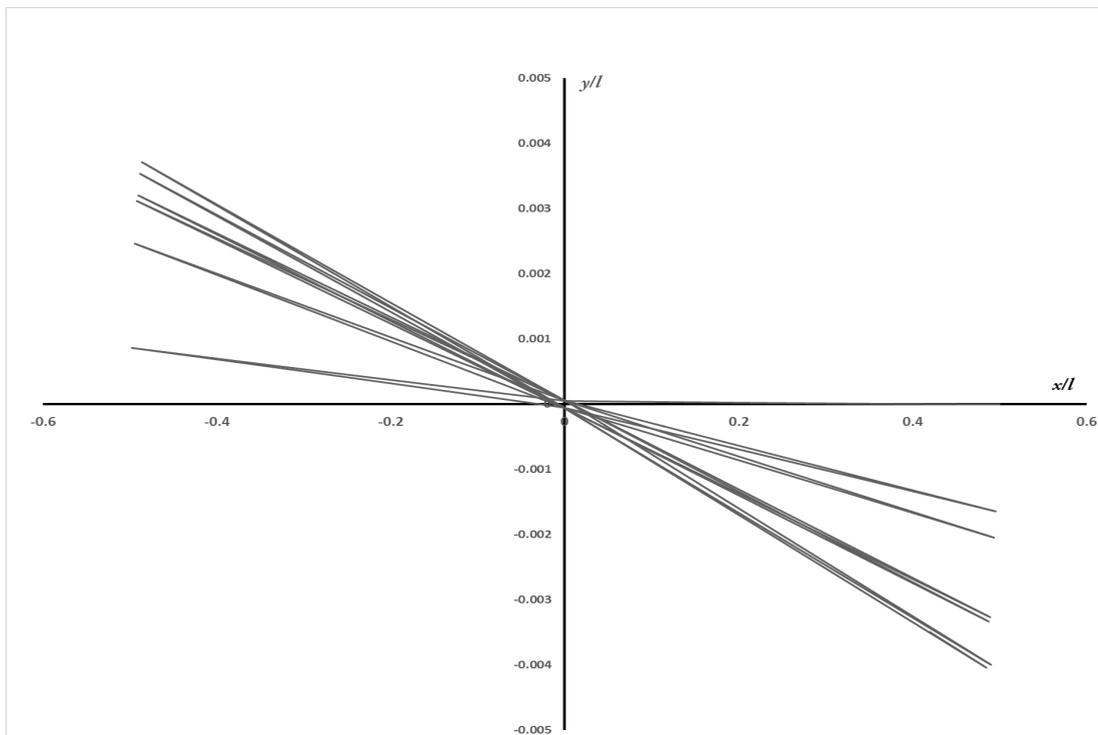

**Figure 6:** The motion trial projected on the horizontal plane for 100 s, derived from the old method (Equation B). Initial parameter: $l = 67m, \alpha_0 = \pi/6, \beta_0 = 0, \dot{\alpha}_0 = 0, \dot{\beta}_0 = 0$.

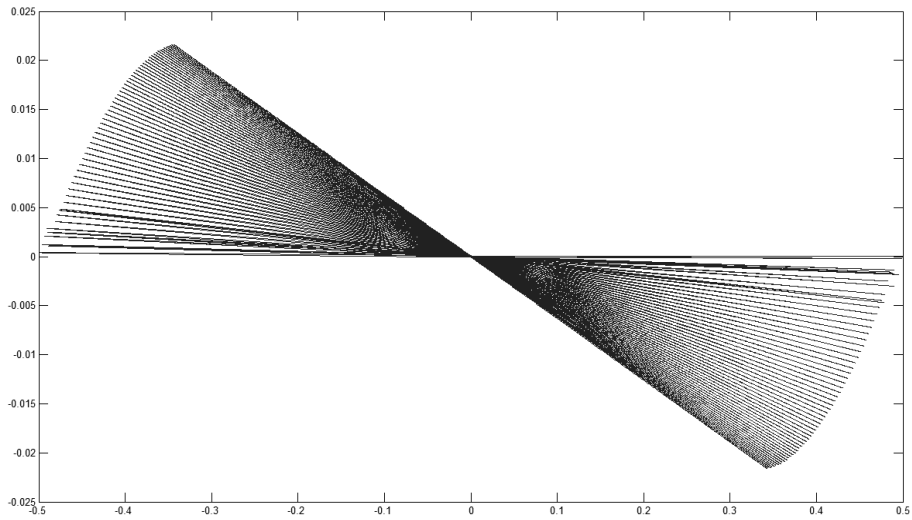

**Figure 7:** The motion trial projected on the horizontal plane for 1000 s, derived from the new method (Equation A). Initial parameter: $l = 67m, \alpha_0 = 0.2rad, \beta_0 = 0, \dot{\alpha}_0 = 0, \dot{\beta}_0 = 0$.

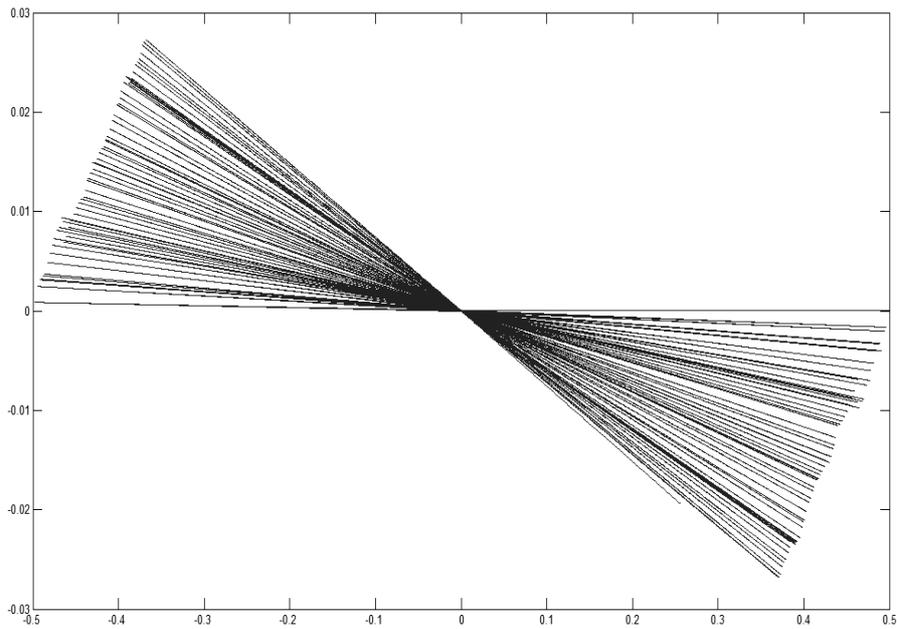

**Figure 8:** The motion trial projected on the horizontal plane for 1000 s, derived from the old method (Equation B). Initial parameter: $l = 67m, \alpha_0 = 0.2rad, \beta_0 = 0, \dot{\alpha}_0 = 0, \dot{\beta}_0 = 0$.

Figure 7 and 8 show that even small oscillation condition is taken, long-time process may lead the solution to the difference. In other words, Equation B is inadequate in long-time process condition.

After that, one element neglected in Equation B but considered in Equation A, the motion in the direction of z-axis will be studied.

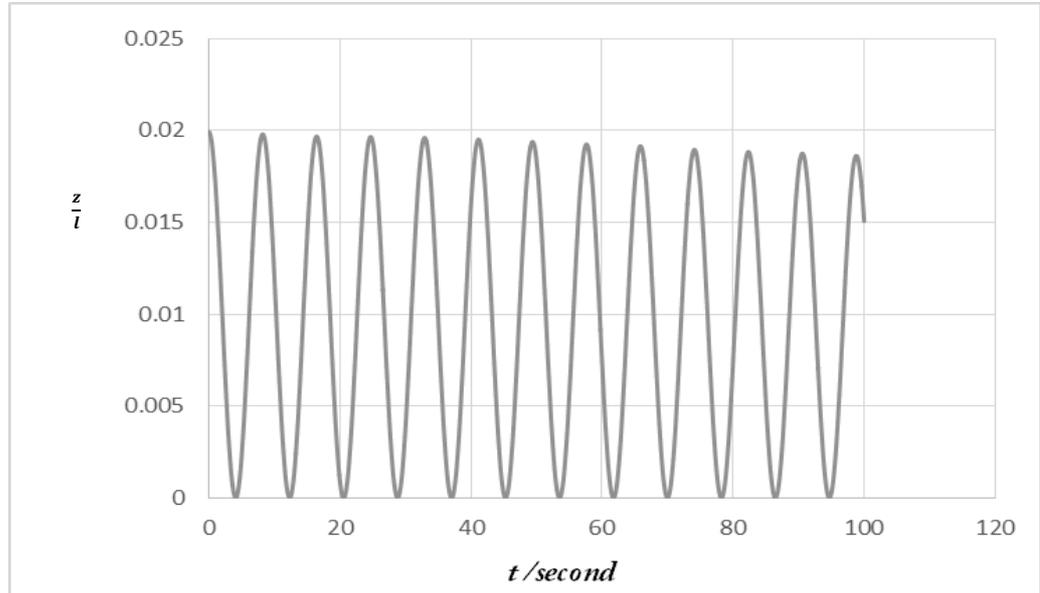

Figure 9: The relationship between z and t. The motion trial projected on the horizontal plane for 1000 s, derived from the new method (Equation A). Initial parameter: $l = 67m, \alpha_0 = 0.2 rad, \beta_0 = 0, \dot{\alpha}_0 = 0, \dot{\beta}_0 = 0$.

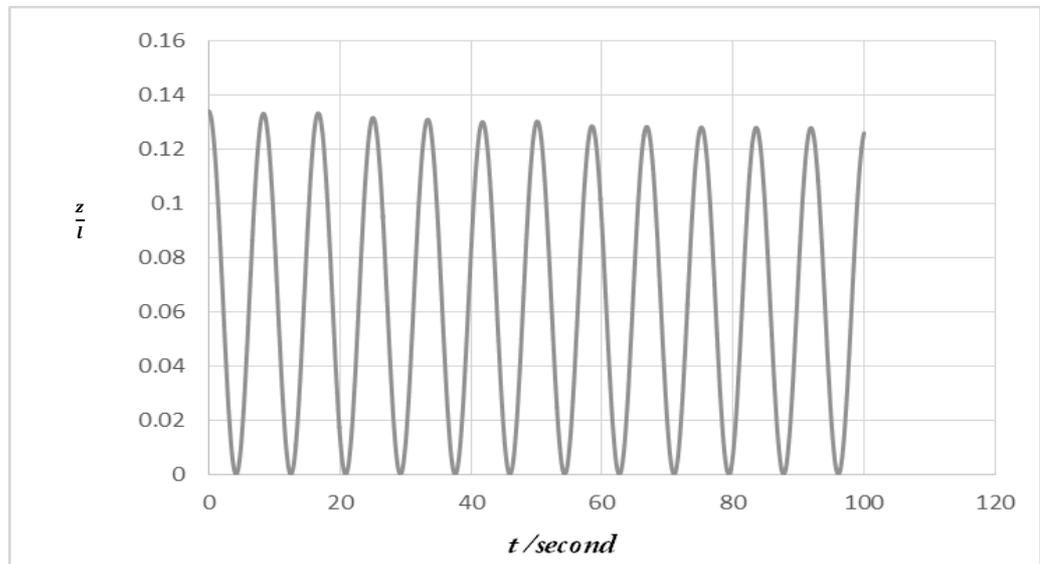

Figure 10: The relationship between z and t. The motion trial projected on the horizontal plane for 1000 s, derived from the new method (Equation A). Initial parameter: $l = 67m, \alpha_0 = \pi/6, \beta_0 = 0, \dot{\alpha}_0 = 0, \dot{\beta}_0 = 0$.

We can know from figure 3-6 and 9-10 that $\frac{z_{max}}{l} = 0.02$, $\frac{x_{max}}{l} = 0.2$ and $\frac{z_{max}}{l} = 0.14$, $\frac{x_{max}}{l} = 0.5$ in low amplitude (0.2 rad) and high amplitude (π/6) respectively, demonstrating that the motion of z-axis direction cannot be neglected especially in the high amplitude condition.

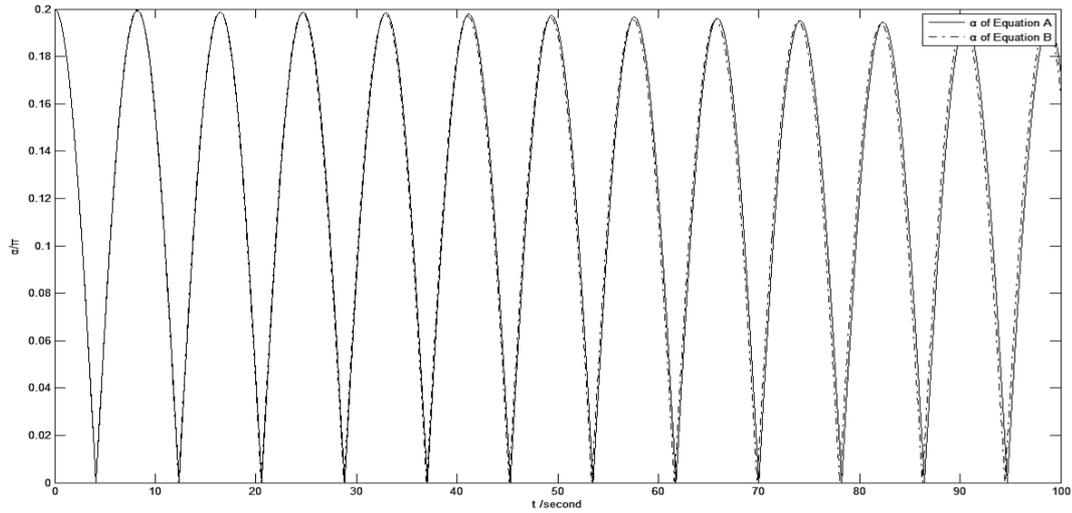

Figure 11: The relationship between $\alpha$ and t for both Equation A and B. (A is the new original method, B is the old method.) Initial parameter: $l = 67m, \alpha_0 = 0.2rad$ $\beta_0 = 0, \dot{\alpha}_0 = 0, \dot{\beta}_0 = 0$. Time span is

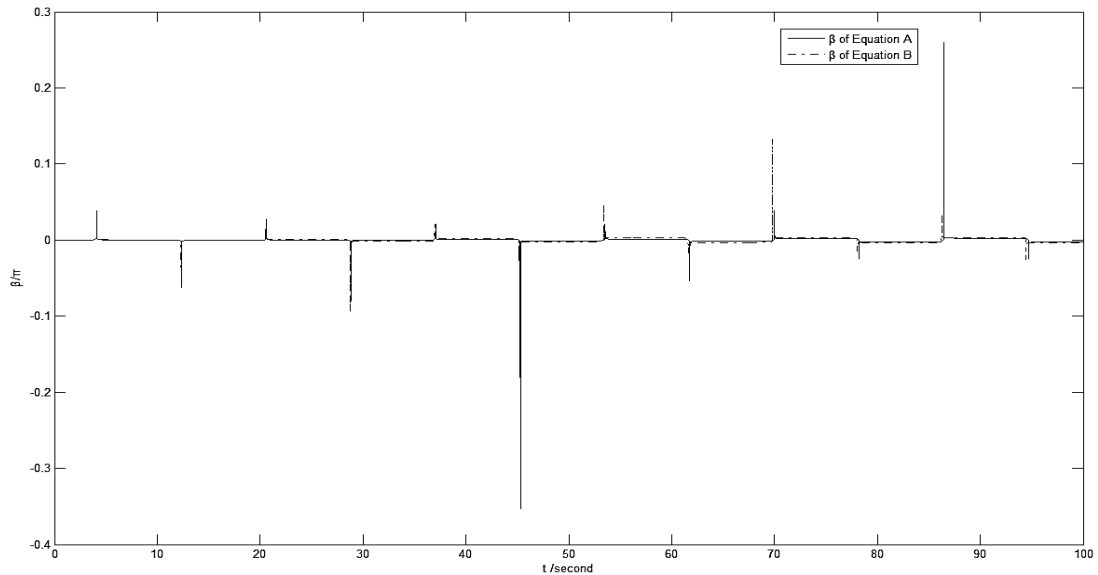

Figure 12: The relationship between $\beta$ and t for both Equation A and B. (A is the new original method, B is the old method.) Initial parameter: $l = 67m, \alpha_0 = 0.2rad, \beta_0 = 0, \dot{\alpha}_0 = 0, \dot{\beta}_0 = 0$. Time span is 100s.

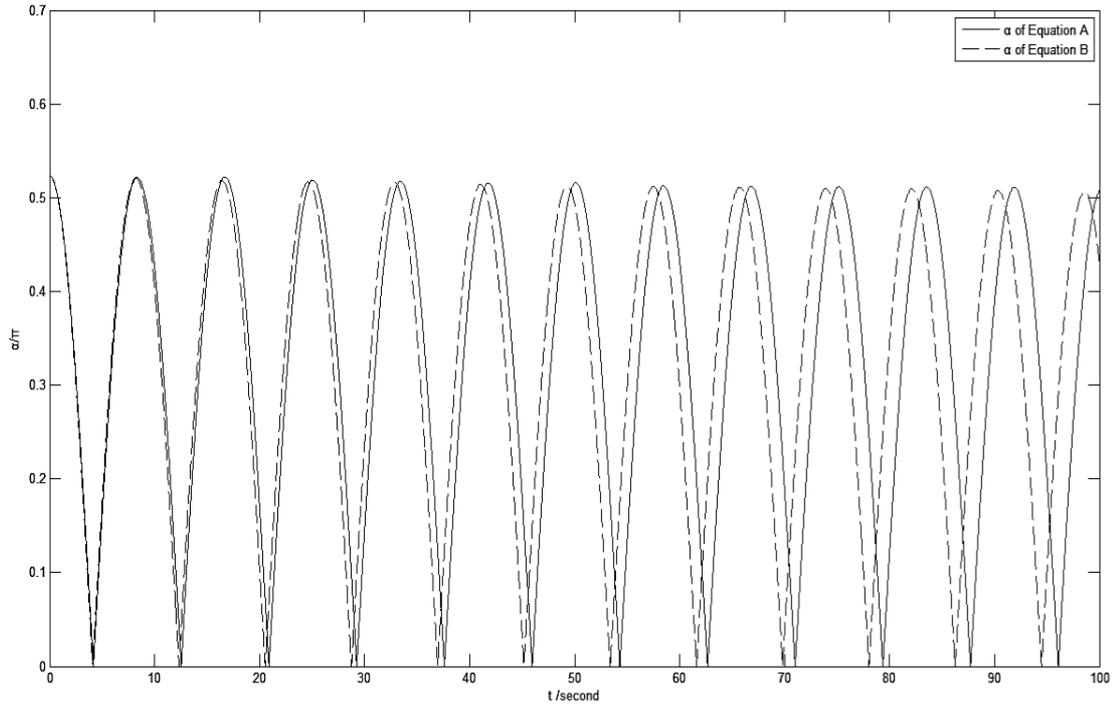

Figure 13: The relationship between $\alpha$ and t for both Equation A and B. (A is the new original method, B is the old method.) Initial parameter: $l = 67m, \alpha_0 = \pi/6, \beta_0 = 0, \dot{\alpha}_0 = 0, \dot{\beta}_0 = 0$. Time span is 100s.

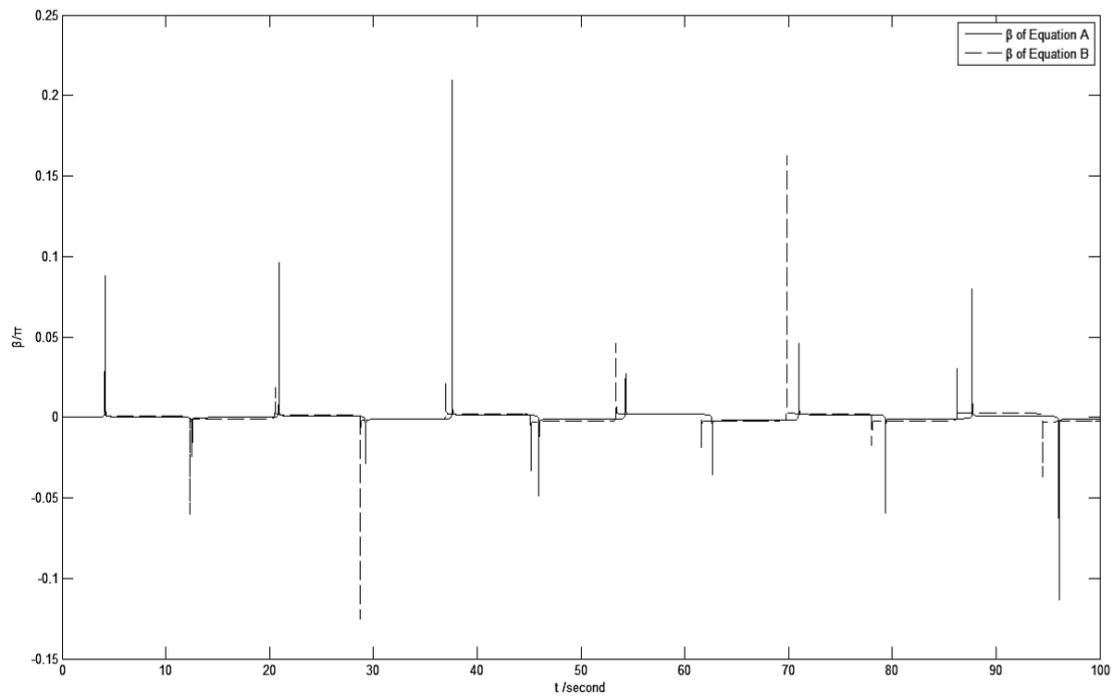

Figure 14: The relationship between $\beta$ and t for both Equation A and B. (A is the new original method, B is the old method.) Initial parameter: $l = 67m, \alpha_0 = \pi/6, \beta_0 = 0, \dot{\alpha}_0 = 0, \dot{\beta}_0 = 0$. Time span is 100s.

Figure 11-14 can also give support for the comparison, similar under small oscillation and different under large, and depict the mode of variation of these variables, near-harmonic for $\alpha$, and near-pulsing for $\beta$.

Finally, there are some more motional features (Figure 15-18) of the Foucault pendulum showing the non-linear feature of it.

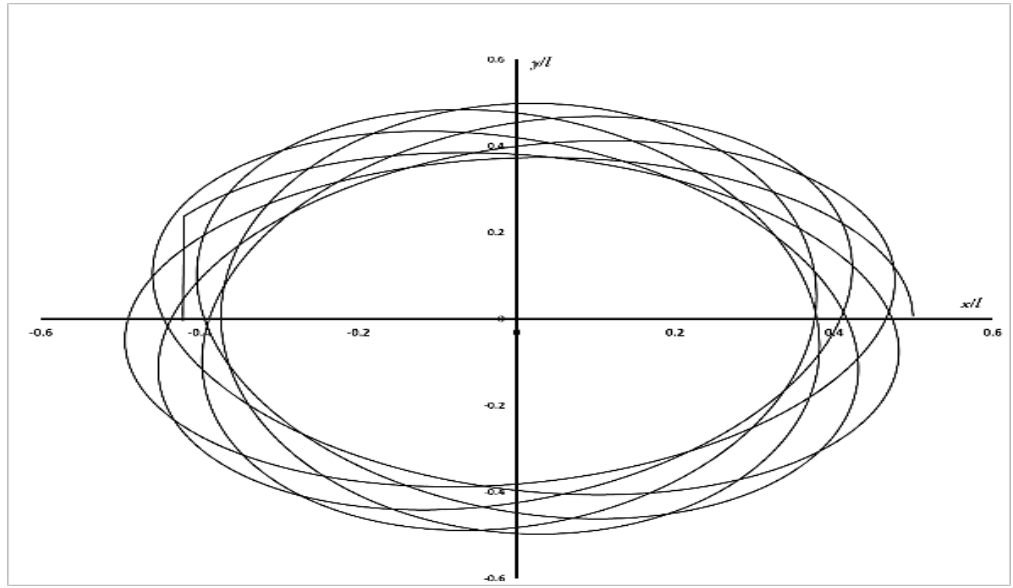

Figure 15: The motion trial projected on the horizontal plane for 100 s, derived from the new original method (Equation A). Initial parameter: $l = 67m, \alpha_0 = \frac{\pi}{6}, \beta_0 = 0, \dot{\alpha}_0 = 0, \dot{\beta}_0 = 0.3$.

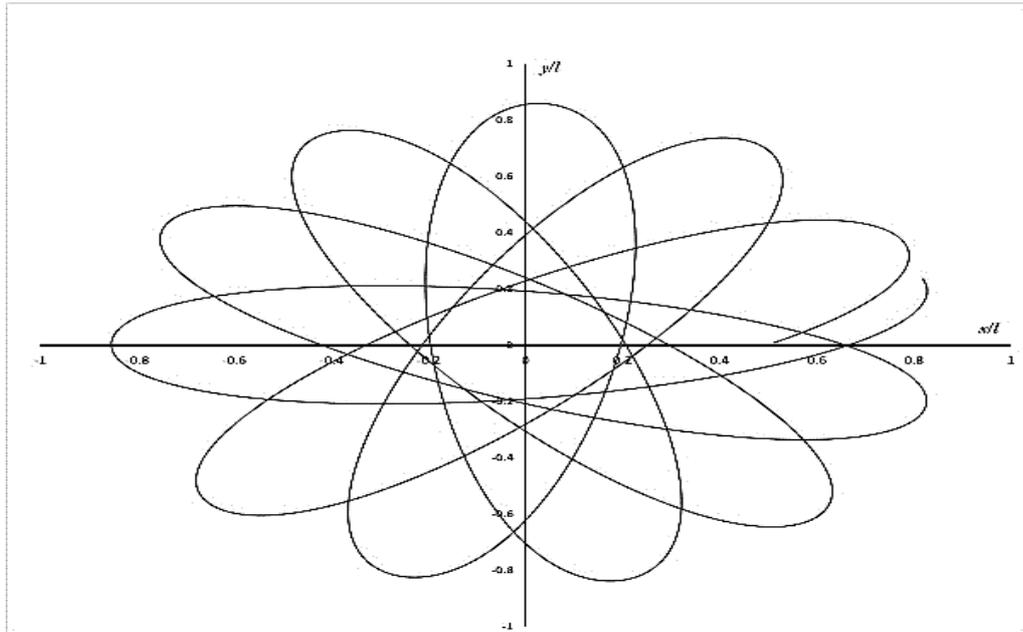

**Figure 16** The motion trial projected on the horizontal plane for 100 s, derived from the new original method (Equation A). Initial parameter: $l = 67m, \alpha_0 = \pi/6, \beta_0 = 0, \dot{\alpha}_0 = 0.3, \dot{\beta}_0 = 0.3$.

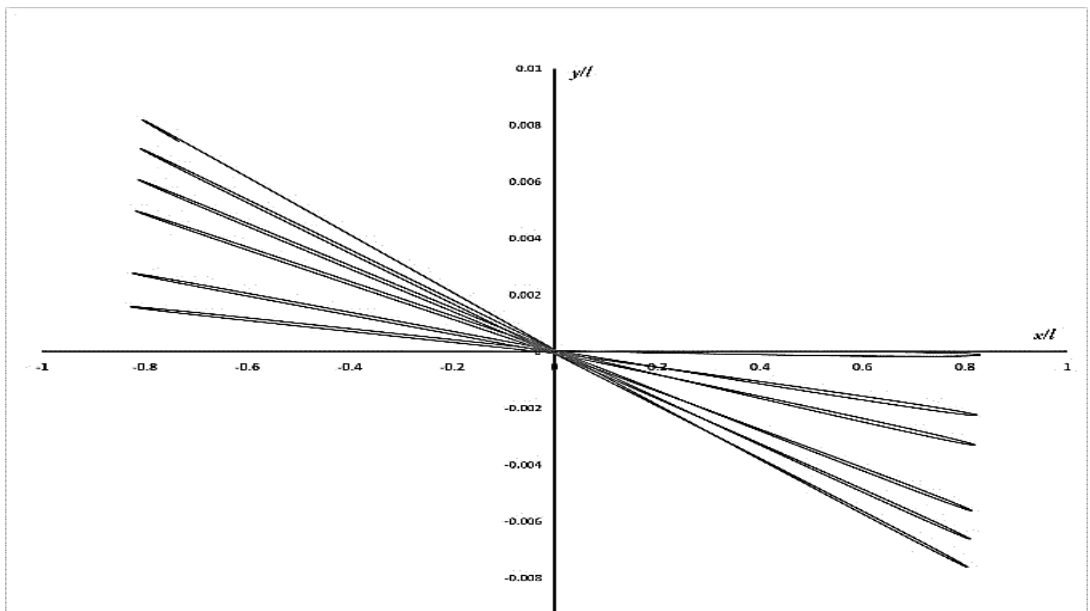

**Figure 17**: The motion trial projected on the horizontal plane for 100 s, derived from the new original method (Equation A). Initial parameter: $l = 67m, \alpha_0 = \pi/6, \beta_0 = 0, \dot{\alpha}_0 = 0.3, \dot{\beta}_0 = 0$.

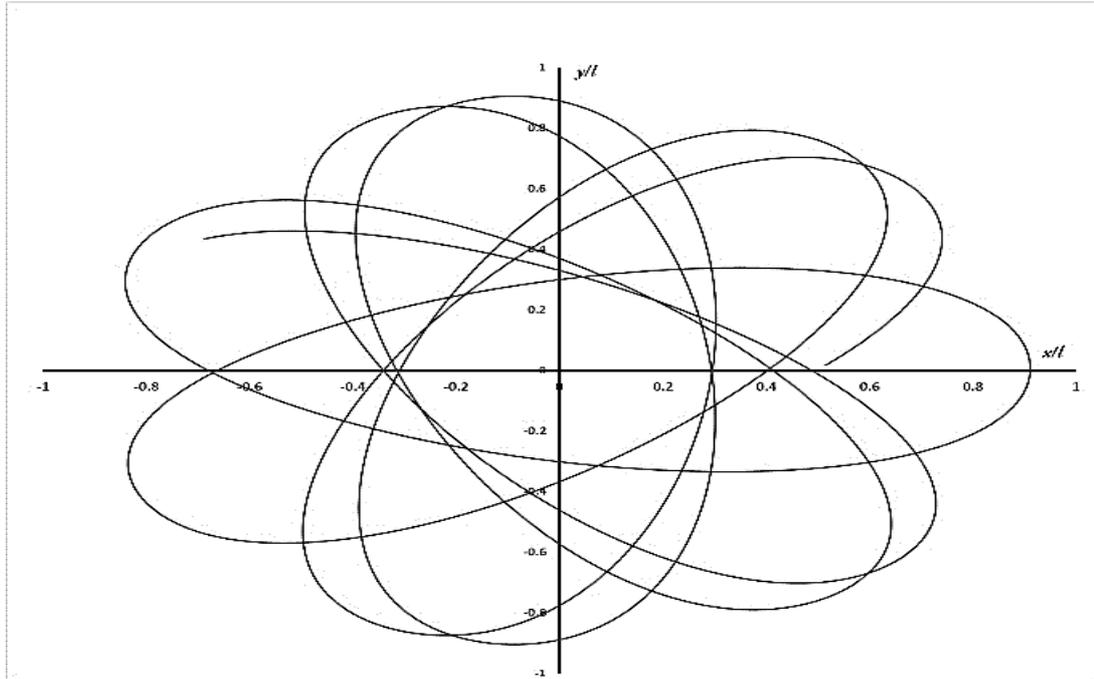

Figure 18: The motion trial projected on the horizontal plane for 100 s, derived from the new original method (Equation A). Initial parameter: $l = 67m, \alpha_0 = \pi/6, \beta_0 = 0, \dot{\alpha}_0 = 0.3, \dot{\beta}_0 = 0.5$

## IV. CONCLUSION

Facing the problem of the Foucault pendulum, we firstly build up the model using some assumptions making the pendulum itself as a simple pendulum except for the applying of the Coriolis force. The De Alembert principle is applied in this problem to avoid calculating the effect of the tension of the rope which is actually do no work to the weight. Then, we take two independent generalized coordinates to describe its exact motion and get the motional equations, marking that the problem is solved theoretically.

During the research, it is found that a large majority of the papers researching on the physic motion of the Foucault pendulum assume that the weight can be seen as moving on the horizontal plane under the condition of small-amplitude oscillation. There is certainly doubt of whether this assumption is considerable and what the motion is like under the condition of high-amplitude oscillation. Thus, we use the model without small-amplitude oscillation and want to compare this two methods to this problem. In this paper, we do a lot of comparison between the two methods in theoretical and numerical analysis. There is a conclusion that the difference between the results are different under varied conditions. Considering the low-amplitude oscillation condition, the two results, to some extent, are similar, but also holds some difference. As for the high-amplitude oscillation, the difference, through the calculation and comparison, is obvious, showing that the original method proposing in this paper holds greater merits than the former.

So as to specialize the motional feature of the Foucault pendulum and do the comparison have mentioned, some numerical simulations are done in this paper. We mainly analyze the motion projected on the horizontal plane, the motion in the vertical direction and the variables (the two generalized coordinates) varying with time. Some features of the non-linear motion are shown in the plots through the numerical calculation.

The problem of the motional feature of the Foucault pendulum is finally solved during the research.